\documentclass[useAMS,usenatbib]{mn2e}
\usepackage[dvips]{graphicx}
\usepackage{amsmath}

\title{A template of atmospheric $O_2$ circularly polarized emission for CMB experiments}
\author[S. Spinelli et al.]{S. Spinelli$^1$,
        G. Fabbian$^{1,2}$,
        A. Tartari$^1$,
        M. Zannoni$^1$
        and M. Gervasi$^1$\\
        $^1$Dipartimento di Fisica ``G. Occhialini'', Universit\`{a} di Milano Bicocca,
        Piazza della Scienza 3, 20126 Milano \\ $^2$ Laboratoire APC - AstroParticule et Cosmologie UMR 7164 CNRS, 10, rue Alice Domon et L\'eonie Duquet 75205 Paris Cedex 13}

\date{\today}
\begin{document}
\maketitle
\begin{abstract}
We compute the circularly polarized signal from atmospheric
molecular oxygen. Polarization of $O_2$ rotational lines is caused
by Zeeman effect in the Earth magnetic field. We evaluate the
circularly polarized emission for various sites suitable for CMB
measurements: South Pole and Dome C (Antarctica), Atacama (Chile)
and Testa Grigia (Italy). An analysis of the polarized signal is
presented and discussed in the framework of future CMB
polarization experiments. We find a typical circularly polarized
signal ($V$ Stokes parameter) of $\sim 50 - 300 \ \mu$K at 90 GHz
looking at the zenith. Among the other sites Atacama shows the
lower polarized signal at the zenith. We present maps of this
signal for the various sites and show typical elevation and
azimuth scans. We find that Dome C presents the lowest gradient in
polarized temperature: $\sim 0.3 \ \mu $K$ /^{\circ}$ at 90 GHz.
We also study the frequency bands of observation: around $\nu
\simeq 100$ GHz and $\nu \simeq 160$ GHz we find the best
conditions because the polarized signal vanishes. Finally we
evaluate the accuracy of the templates and the signal variability
in relation with the knowledge and the variability of the Earth
magnetic field and the atmospheric parameters.

\end{abstract}

\section{Introduction}
The observation of the Cosmic Microwave Background (CMB) from the
ground can be performed inside the atmospheric windows, far from
the main emission lines of the most abundant molecules. For the
experiments devoted to the measurement of the polarization
properties of the CMB it is mandatory to evaluate, in addition,
the effects of the atmosphere. In a recent paper
\citep{pietranera} the effects of ice crystal clouds in the upper
troposphere have been considered, but the major effect arises from
the polarized emission of the $O_{2}$ molecule. The relevance of
this effect for CMB measurements has been pointed out by
\cite{hanany}, where for the first time a quantitative evaluation
of the polarized signals has been presented. Also \cite{keating}
discuss the relevance of the polarized emission of the $O_{2}$ as
a foreground for CMB observations.

Here we compute the polarized emission of the $O_{2}$ molecule,
developing the formalism introduced in the pioneering works by
\cite{len2,len1}. The interaction of the Earth magnetic field with
the magnetic dipole moment of the atmospheric $O_{2}$ molecule
induces Zeeman splitting onto the roto-vibrational lines. As a
result of this interaction, radiation becomes weakly circularly
polarized with a degree of polarization $\sim10^{-4} - 10^{-5}$. A
linear polarization is also generated, but the signal is three or
four orders of magnitude lower and still below the current
detection sensitivity.

This circularly polarized signal can be considered a contaminant for
CMB ground-based experiments for two reasons. The first one is that
non-idealities in the instrument can produce conversion from
circular to linear polarization: this leakage from the $V$ Stokes'
parameter to $\{ Q,U\}$ can be the source of a spurious signal if
the circularly polarized radiation of atmospheric oxygen is
picked-up by the instrument. Here the effect on the recovery of the
cosmological polarized signal is evident since CMB is expected to be
only (weakly) linearly polarized \citep{hu}.

The second one is that there is an increasing theoretical effort
dedicated to the physical mechanisms able to create a non
vanishing CMB circular polarization. In fact, there are several
attempts to add terms to the standard electromagnetic lagrangian
density, and to predict observable effects arising from these
extensions. Among others, and without sake of completeness, we
recall: test of the Shiff conjecture on the EEP \citep{carroll};
coupling between photons and slowly varying pseudoscalar fields
\citep{harari}, \citep{carlson} and \citep{finelli}; coupling
between photons and pseudoscalar fields in presence of
intergalactic magnetic fields \citep{agarwal}; coupling between
photons and vector fields to test Lorentz-invariance violating
processes \citep{alexander}. Authors, in all of these cases, show
how a fundamental process induces optical activity in the
Intergalactic Medium. In fact, the extensions of the standard
lagrangian generate new terms in Maxwell's equations (see eq. 3 in
\cite{harari}), therefore leading to modified dispersion
relations. We have to mention also the possibility of having a
$V$-mode induced by last-scattering in a magnetized primordial
plasma (see \cite{giovannini I,giovannini II}) and, after
recombination, by intergalactic magnetic fields through Faraday
conversion \citep{cooray}.

It is likely that, with increasing theoretical motivations, a
dedicated experiment will be proposed in the near future to update
the very early upper limits obtained by \cite{lubin} and
\cite{partridge}. The present work has been prepared also in this
spirit.

In section 2 the basic physics of the process is presented, while
in section 3 the tensor radiative transfer approach used to
compute the atmospheric integrated signal is described. The
analysis is applied to different sites: Dome C (Antarctica), South
Pole, Atacama (Chile), Testa Grigia (Italian Alps), where mm-wave
instruments are located. Results are reported in section 4. We
compare the effect in different places and define the optimal
observation bands in order to minimize the effect from this
contaminant.

\section{Polarized signal theory}
$O_2$ is the only abundant atmospheric molecule with non-zero
magnetic dipole moment, because the two electrons in the highest
energy state couple with parallel spin. The presence of polarized
radiation in the atmosphere is associated with the interaction of
the magnetic dipole moment of the $O_2$ molecule and the Earth
magnetic field. In the millimeter wave region of the
electromagnetic spectrum, $O_2$ molecule presents a set of lines
due to roto-vibrational transitions. The presence of the Earth
magnetic field induces Zeeman splitting on these lines, which then
become polarized.

Because of the Zeeman effect a quantum state represented by the
quantum number $J$ is split in $2J+1$ states, each of them described
by a different value of $M_J$, under the condition $|M_J|\leq J$.
The selection rules permit the following transitions, $\Delta J = +
1$ (called $L^+$) and $\Delta J = - 1$ (called $L^-$):

\begin{eqnarray}
\begin{array}{llllll}
& \Delta J = + 1 \ : \ & \ J = L & \ \rightarrow & \ J = L+1 & \ \
(L^+) \\ \nonumber & \Delta J = - 1 \ : \ & \ J = L & \
\rightarrow & \ J = L-1 & \ \
 (L^-)
\end{array}
\end{eqnarray}

\noindent Under these conditions, selection rules permit to have
$\Delta M_J=0$ ($\pi$ transitions) and $\Delta M_J=\pm1$
($\sigma_{\pm}$ transitions). In Table \ref{tab_freq2} we report
the frequencies of $O_2$ roto-vibrational transitions allowed by
the {\it Fermi Golden Rule} and used for our calculation, together
with their corresponding quantum number ($L^{\pm}$). We stop to
$L^{\pm} = 37$ because of lack of information about the line
broadening and mixing parameters (see \cite{Liebe}) of higher
orders. We also verified the effect of adding lines with $L^{\pm}
> 37$: far from the transition frequencies the difference is
negligible ($\Delta V$/$V < 10^{-5}$).

\begin{table}
\begin{center}
\caption{Line frequency and corresponding quantum number
$L^{\pm}$ used for the computation of the atmospheric $O_2$
polarized signal.}\label{tab_freq2}
\begin{tabular}{rrr}
  \hline
  $L^{\pm}$ & $\nu_t(L^-)$ \ (GHz) & $\nu_t(L^+)$  \ (GHz) \\
  \hline
\textbf{1}  & 118.750343  & 56.264777 \\
\textbf{3}  &  62.486255  & 58.446580 \\
\textbf{5}  &  60.306044  & 59.590978 \\
\textbf{7}  &  59.164215  & 60.434776 \\
\textbf{9}  &  58.323885  & 61.150570 \\
\textbf{11} &  57.612480  & 61.800155 \\
\textbf{13} &  56.968180  & 62.411223 \\
\textbf{15} &  56.363393  & 62.997977 \\
\textbf{17} &  55.783819  & 63.568520 \\
\textbf{19} &  55.221372  & 64.127777 \\
\textbf{21} &  54.671145  & 64.678898 \\
\textbf{23} &  54.130002  & 65.224065 \\
\textbf{25} &  53.595751  & 65.764744 \\
\textbf{27} &  53.066908  & 66.302082 \\
\textbf{29} &  52.542392  & 66.836820 \\
\textbf{31} &  52.021405  & 67.369589 \\
\textbf{33} &  51.503339  & 67.900867 \\
\textbf{35} &  50.987728  & 68.431005 \\
\textbf{37} &  50.474204  & 68.960312 \\
\hline
\end{tabular}
\end{center}
\end{table}

Radiation intensity and polarization direction are also determined
by the angle ($\theta$) between the line of sight and the Earth
magnetic field. We must define a reference system onto which the
vector $\bmath{E}$ of the electromagnetic wave is projected. A
good choice is the 3D geomagnetic system (see \cite{len1})
centered at the geographic Earth center: $z$ axis correspondes to
the geomagnetic dipole axis and angles $\psi_m$ and $\phi_m$
identify the geomagnetic latitude and longitude respectively.
Matrix representing the polarized intensity \textbf{I} can be
evaluated through the coherence matrix ($\brho$) for the different
transitions:

\begin{eqnarray}
&&\textbf{I}_{\Delta M_J} (S,L,M_J) = P_{trans,\Delta M_J}(S,L,M_J)
\ \brho_{\Delta M_J}, \nonumber
\end{eqnarray}

\noindent In this reference frame the matrices for the three
different transitions become:

\begin{eqnarray}
&&\Delta M_J=-1 \ : \ \ \ \ \brho_{\sigma_{-}} =
\left(
  \begin{array}{cc}
    1 & i\cos\theta \\
    -i\cos\theta & \cos^{2}\theta \\
  \end{array}
\right)\label{rho-}\\
&&\Delta M_J=+1 \ : \ \ \ \ \brho_{\sigma_{+}} =
\left(
  \begin{array}{cc}
    1 & -i\cos\theta \\
    i\cos\theta & \cos^{2}\theta \\
  \end{array}
\right)\label{rho+}\\
&&\Delta M_J=0 \ : \ \ \ \ \ \ \brho_{\pi} \ \ =
\left(
  \begin{array}{cc}
    0 & 0 \\
    0 & \sin^{2}\theta \\
  \end{array}
\right)\label{rho0},
\end{eqnarray}

\noindent where $P_{trans}$ denotes the transition probability for
the line identified by the quantum numbers $(S,L,J,M_J)$ and fix
their relative intensity. Equations (\ref{rho-}) - (\ref{rho0})
show that $\sigma_-$ and $\sigma_+$ lines produce circular
polarization, due to the off-diagonal terms in the matrices.
Besides the combination of the three transitions generate also a
small fraction of linear polarization.

The total coherence matrix $\mathbfss{A}_{tot}$ can be derived
by expressing the whole contribution given by the superposition of
the different lines:

\begin{eqnarray}
\mathbfss{A}_{tot}=\! C(\nu,P,T)\!\!\!\!\sum_{\Delta
M_J=-1}^{\Delta M_J=+1}\!\!\!\brho_{\Delta
M_J}\!\!\!\!\!\!\sum_{M_{J}=-J}^{+J}\!\!\!\!\!\!\!\!P_{trans}(S,L,M_J,\Delta
J,\Delta
M_J) \nonumber\\
F(\nu,\nu_k,\Delta \nu_c).\nonumber
\end{eqnarray}

\noindent \textbf{I} matrices are multiplied by two weight factors:
the first one, $C(\nu,P,T)$, takes into account the dependence on
pressure, temperature and frequency \citep{len1}:

\begin{equation} \label{coeff}
C(\nu,T,P)=0.229\frac{P\nu^2}{T^3}e^{\frac{-E_L}{K_B T}};
\end{equation}

\noindent the second function, $F(\nu,\nu_k,\Delta\nu_c)$, is the
line profile. We neglect the Doppler effect and consider only the
collisional line broadening, which is dominant in the lower layers
of the atmosphere where most of the signal is generated. Therefore
we use the collisional line profile \citep{VV} corrected for the
line mixing effects as shown by \cite{Rosen_Staelin} and
\cite{Liebe}. The line profile used is then:

\begin{equation}\label{profile}
F(\nu, \nu_k, \Delta \nu_c) = F_{\Delta} - F_{Y}
\end{equation}

\noindent where:

\begin{eqnarray}
F_{\Delta} = \Delta \nu_c \left[ \frac{1}{(\nu_k - \nu)^2 + \Delta
\nu_c^2} + \frac{1}{(\nu_k + \nu)^2 + \Delta \nu_c^2} \right] \nonumber \\
F_{Y} = Y_k \left[ \frac{\nu_k - \nu}{(\nu_k - \nu)^2 + \Delta
\nu_c^2} + \frac{\nu_k + \nu}{(\nu_k + \nu)^2 + \Delta \nu_c^2}
\right] \ . \nonumber
\end{eqnarray}

\noindent Here the line broadening is:

\begin{equation}
\Delta \nu_c = A_0 \ P \left(\frac{300}{T}\right)^{0.8}
\end{equation}

\noindent and the line mixing coefficient is:

\begin{equation}
Y_k = \left(A_1 + A_2 \frac{300}{T} \right) P
\left(\frac{300}{T}\right)^{0.8} \ .
\end{equation}

\noindent Here the coefficients $A_0$, $A_1$ and $A_2$, for each
transition, are given in table 2 of \cite{Liebe}, where
calculations are compared with measurements. In eq. \ref{profile}
$\nu_k = \nu_t + \delta_{\Delta M_J}$, where $\nu_t$ is the
transition frequency, shown in table \ref{tab_freq2} and
$\delta_{\Delta M_J}$ is a Zeeman induced frequency shift
\citep{len1} given a magnetic field $\bmath{B}$ (nT):

\begin{displaymath}
\delta_{\Delta M_J}(MHz)=\left\{ \begin{array}{ll}
\frac{K_0 B}{L+1}\left( \Delta M_J+M_J\displaystyle\frac{L-1}{L} \right) & \ \ (L^+) \\
- \frac{K_0 B}{L}\left( \Delta M_J+M_J\displaystyle\frac{L+2}{L+1}
\right) & \ \ (L^-)
\end{array}\right.
\end{displaymath}

\noindent Here $K_0 = 2.8026 \times 10^{-5}$ and $\Delta M_J = 0,
\pm 1$. Taking into account all these relations and summarizing
them, the total coherence matrix can be written as:

\begin{eqnarray}
\mathbfss{A}_{tot} = a \brho_{\sigma_-} + b \brho_{\sigma_+} +
c \brho_{\pi} = \nonumber \\
\left(\begin{array}{cc}
  a+b & i(a-b)\cos\theta \\
  -i(a-b)\cos\theta & (a+b) \cos^2 \theta + c \sin^2 \theta \\
  \end{array}\right) = \nonumber \\
 \left(\begin{array}{cc}
  \alpha & i \beta\\
   -i \beta& \gamma
   \end{array}\right)
\end{eqnarray}

\noindent This matrix expresses the coherence (polarization)
properties of the signal coming from the Zeeman split $O_2$ lines.
We get a matrix for a given atmospheric condition of pressure,
temperature and height. In particular we notice that: \textbf{1)}
off-diagonal terms are non vanishing and conjugate purely
imaginary complex numbers thus radiation is only circularly
polarized ($V \neq 0$) since $U$ parameter is null; \textbf{2)}
circular polarization is larger when the line of sight is aligned
with the Earth magnetic field: $V \propto \beta = (a-b)
\cos\theta$; \textbf{3)} the diagonal terms are still different
($\alpha \simeq \gamma$), so that Stokes' parameter $Q \propto
\alpha - \gamma \neq 0$; \textbf{4)} linear polarization is larger
when the line of sight is orthogonal to the Earth magnetic field:
$Q \propto (a+b-c) \sin^2\theta$; \textbf{5)} the component of non
vanishing linear polarization ($Q$ parameter) is nearly aligned
with the Earth magnetic field, even if a small Faraday rotation
may occur (see \cite{Rosen_Staelin}). Actually it can be seen that
the ratio of linear ($Q$) to circular ($V$) polarization is $Q/V
\sim 10^{-4}$ (see \cite{hanany}). In the present paper we
evaluate only the circularly polarized signal, as the linearly
polarized one is still below the sensitivity of the current
experiments (see section 4).

\section{Coherence brightness temperature}
\subsection{Modeling the atmosphere}
The propagation of radiation emitted by Oxygen molecules
throughout atmosphere modifies its polarization properties. We
compute the $\mathbfss{A}$ matrix via a tensor radiative transfer
approach. We compute the total coherence matrix in terms of
brightness temperature and divide the atmosphere in $N$ layers of
thickness $\Delta z = 0.2$ km. We interpolate the atmospheric
parameters profiles available in literature to fit this thickness.
Than we take the average value of the physical parameters inside
each layer and use the theory of radiative transfer, described in
\cite{len1}, to compute the signal emerging from one layer, taking
into account all the layers lying above it:

\begin{equation}
\mathbfss{T}_B(z,\nu)=\sum_{i=1}^{N}\mathbfss{WF}(\Delta
z_i,\nu)T_i.
\end{equation}
\noindent $\mathbfss{WF}$ is a weight function for the physical
temperature $T_i$ of the $i$-th layer, in which all the
information about the transfer properties of the layer are
encoded, according to the formalism developed in \cite{len2}:

\begin{equation}
\mathbfss{WF}(\Delta z_i,\nu) = \mathbfss{P}(\Delta z_i,\nu)
\left[\mathbfss{I}-e^{-2\mathbfss{A}_i(\nu) \Delta z_i}\right]
\mathbfss{P}^*(\Delta z_i,\nu).
\end{equation}

\noindent In this relation $\mathbfss{A}_i$ expresses the total
attenuation matrix of the i-th layer (see section 2) and
$\mathbfss{P}$ is the product of the exponential matrices taking
into account the absorption of radiation through different layers,
that is $\mathbfss{P}=\mathbfss{I}$ for $i=1$,
$\displaystyle\mathbfss{P}=e^{-\mathbfss{A}(\nu)\Delta z_i}$ for
$i=2$ while for $i>2$ we have:

\begin{equation}
\mathbfss{P}(\Delta z_i,\nu)=e^{-\mathbfss{A}_1(\nu) \Delta
z_i}\times .... \times e^{-\mathbfss{A}_{i-1}(\nu) \Delta z_i}.
\end{equation}

\noindent To quantify the brightness temperature matrix, we have
developed a code in which we change the input parameters of the
location (geographic co-ordinates) and of the instrument (pointing
directions) considered. The brightness temperature is therefore
computed including the contribution of all the transitions of the
oxygen molecule, ranging from $\sim$ 50 GHz to 118.7 GHz shown in
Table \ref{tab_freq2}.

\subsection{Atmospheric parameters and Magnetic Field}
Vertical profiles for atmospheric pressure and temperature have
been taken from \cite{tomasi} for Dome C station, from
\cite{depetris} for Testa Grigia, while for
Atacama\footnote{http:$//$www.tuc.nrao.edu/alma/site/Chajnantor/instruments\\/radiosonde/}
and South
Pole\footnote{ftp:$//$ftp.cmdl.noaa.gov/ozwv/ozone/Spole/} sites
atmospheric parameters are available on their related web sites.

The Earth magnetic field has been calculated with the NASA tool
based on the IGRF-2010
model\footnote{http:$//$omniweb.gsfc.nasa.gov/vitmo/cgm\_vitmo.html}.
IGRF is a model of the geomagnetic field close to Earth surface,
generated inside the Earth
body\footnote{http:$//$www.ngdc.noaa.gov/IAGA/vmod/igrfhw.html}.
The field can be roughly represented as a dipole. Its intensity on
the Earth surface is $\sim 3 - 6 \times 10^4$ nT. IGRF describes
the lower spatial frequencies of the field: IGRF-2010 is truncated
at the multipole $n=13$, corresponding to an angular scale of
$\sim$ 15$^{\circ}$ and to a typical wavelength of $\sim 3000$ km along
the surface. Higher spatial frequencies, which represent a small
scale correction to the model, are not accounted for. These
components come mainly from the magnetized rocks of the crust and
are typically $\sim$ 5-10 nT rms (see \cite{Hulot}). An additional
contribution comes from the external ring currents present in the
ionosphere and magnetosphere. The external field intensity, in
quiet conditions of the solar activity, is only a few tens of nT.
Earth magnetic field is also changing with time. The internal
field is slowly changing with a global rate of $\sim 80$ nT/year.
The external field is more rapidly changing and increases up to 2
orders of magnitude during magnetic storms. Uncertainty of
IGRF-2010 model, in a typical quiet condition, for the current
epoch is estimated to be $\sim 10$ nT rms for the total intensity
and $\sim 1$ arcmin for the direction (see IGRF "health" warning,
by F.J. Lowes$^4$, IAGA Working Group VMOD, January 2010).

\begin{figure}
\begin{center}
\includegraphics[width=75mm]{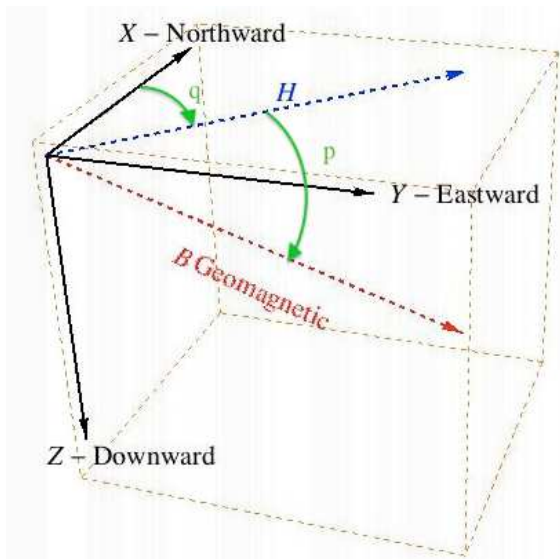}
\caption{\label{ref_B} Reference system used to describe the
magnetic field ($\bmath{B}$) direction: $\bmath{H}$ is the
projection of $\bmath{B}$ onto the horizontal plane. The angles
$\hat{q}$ and $\hat{p}$ define the direction with respect to the
geographic North (X axis) and the inclination with respect to the
local horizontal (orthogonal to Z axis). Downward directions mean
positive values.}
\end{center}
\end{figure}

Data concerning geographic, atmospheric and geomagnetic parameters
of the different observing sites are reported in Table
\ref{tab_data}. Typical atmospheric data and magnetic field are
evaluated on the ground. In Figure \ref{ref_B} a picture of the
local reference frame used to describe the magnetic field
$\bmath{B}$ direction is shown: $\hat{p}$ is the angle between
$\bmath{B}$ and the local horizontal; $\hat{q}$ is the angle
between the horizontal projection ($\bmath{H}$) and the local
North. For South Pole X axes is pointing towards the geographic
longitude $\ell = 0$.

\begin{table*}
\caption{\label{tab_data}Geographic and atmospheric data on the
ground for the various sites considered. For a description of the
magnetic field direction see Fig. \ref{ref_B}. Magnetic field
parameters are evaluated for 2010.}
\begin{tabular}{|c|c|c|c|c|c|c|c|c|}
  \hline
  Site & Latitude ($^{\circ}$) & Longitude ($^{\circ}$) & h (km) & P (mb) & T (K) &
  B (nT) & $\hat{p}$ ($^{\circ}$) & $\hat{q}$ ($^{\circ}$) \\
  \hline
  Dome C       & 75.06 S & 123.21 E & 3.233 & 630 & 230 & 62713 & -80.8 & -140.4 \\
  South Pole   & 90.00 S &        0 & 2.800 & 680 & 242 & 55849 & -72.8 &  -28.2 \\
  Atacama      & 23.01 S &  67.45 W & 5.104 & 556 & 270 & 23296 & -20.2 &   -6.3 \\
  Testa Grigia & 45.56 N &   7.48 E & 3.480 & 655 & 260 & 46828 &  61.9 &    0.7 \\
  \hline
\end{tabular}
\end{table*}

\section{Results}
\subsection{Frequency templates}
Using the procedure and the code described in the previous
sections we have been able to evaluate the polarized signal of the
atmospheric oxygen for each place on the Earth surface, assuming
that the atmospheric parameters (basically pressure and
temperature profiles) are available. Earth magnetic field
parameters are computed for the year 2010.

\begin{figure}
\begin{center}
\includegraphics[width=75mm]{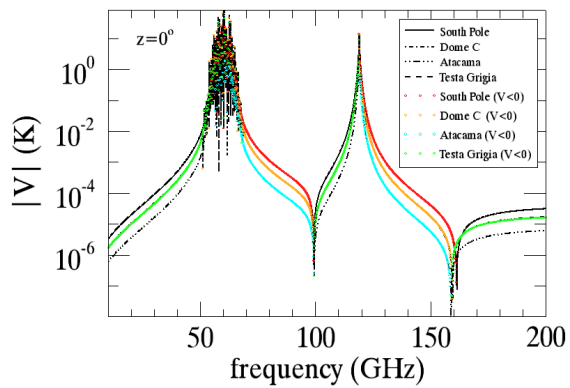}
\caption{Polarized atmospheric $O_2$ signal emitted at the zenith
for the various observing sites as a function of the frequency.
The absolute value of $V$ is shown. Colored lines denote negative
values. Values obtained at Dome C are very similar to those found at Testa Grigia,
although the sign is the opposite.} \label{pol_comparison}
\end{center}
\end{figure}

We computed the circularly polarized contribution ($V$ Stokes
parameter) for the various sites previously mentioned. The
absolute value of the signal at the zenith {\it vs} frequency is
shown in Figure \ref{pol_comparison}. The strong contribution of
the single line at 118 GHz and of the ensemble of lines around 60
GHz can be immediately noticed.

Differences among the various sites are due to a number of
reasons. Among them the most important are: \textbf{1.} the
altitude of the observing site since the lower layers of the
atmosphere, having an higher pressure (see eq. \ref{coeff}), are
the major contributors to the signal; \textbf{2.} at mid and low
latitudes (Atacama and Testa Grigia) the amplitude of the Earth
magnetic field is a factor of two weaker than that at South Pole;
\textbf{3.} the amplitude of the effect depends on the scalar
product between the line of sight and the Earth magnetic field
direction. For sites close to the magnetic poles the magnetic
field direction is almost vertical, while at lower latitudes the
magnetic field direction is almost horizontal and aligned along
the North-South direction.

Black lines in Figure \ref{pol_comparison} denote values $V>0$ (i.e. right handed circular
polarization), while colored lines denote values $V<0$ (i.e. left
handed circular polarization). We can notice that the polarized
signal at Testa Grigia has a sign reversed respect to other
places. This happens because the magnetic field \textbf{B} is
directed downward in Testa Grigia (which is in the northern
hemisphere), while \textbf{B} is directed upward elsewhere (all
the other sites are located in the southern hemisphere). We can
notice also a sign reversal at the frequencies $\nu \simeq 99$
GHz and $\nu \simeq 160$ GHz. The first null is originated by the
118 GHz line wing crossing the 50-70 GHz lines wings. The null at
higher frequency is originated adding the contribution from lines
with $L \geq 9$. Adding lines with higher $L$ the null frequency
shifts to lower values, converging to the final value shown in
Fig. \ref{pol_comparison}. The coefficients $A_0$, $A_1$ and $A_2$
(see eq. 6 and 7), which assume different values line by line,
seem to play an important role in this effect. Anyway, as discussed in section \ref{acc_temp},
it should be stressed that above $\sim$120 GHz the model is not yet validated, and calculation
could be not accurate.

\subsection{Angular templates}
\begin{figure}
\begin{center}
\includegraphics[width=80mm]{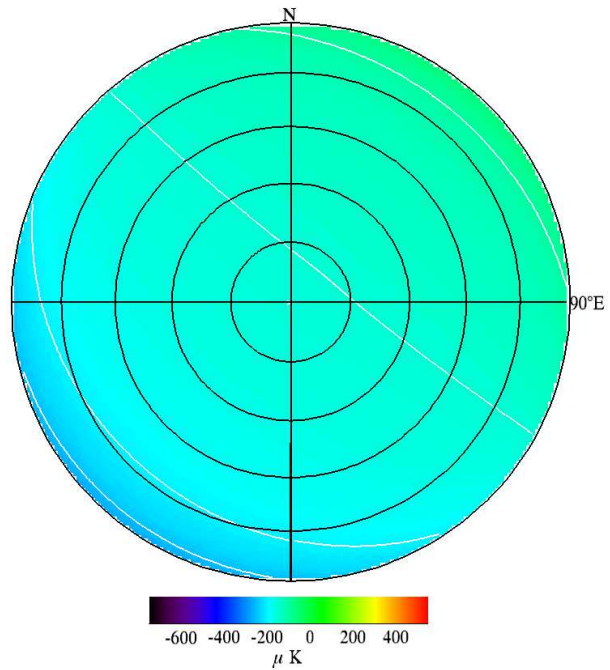}
\caption{Map of the atmospheric $O_2$ circularly polarized signal
at Dome C, at frequency $\nu = 90$ GHz. The Lambert projection
centered at the local zenith, down to 15$^{\circ}$ of elevation, is
shown: upward direction is the geographic North. Black circles
denote elevation intervals of 15$^{\circ}$. Color scale ranges from
-750 $\mu K$ to +550 $\mu K$. White contours denote intervals of
50 $\mu K$ in brightness temperature.} \label{dc}
\end{center}
\end{figure}

\begin{figure}
\begin{center}
\includegraphics[width=80mm]{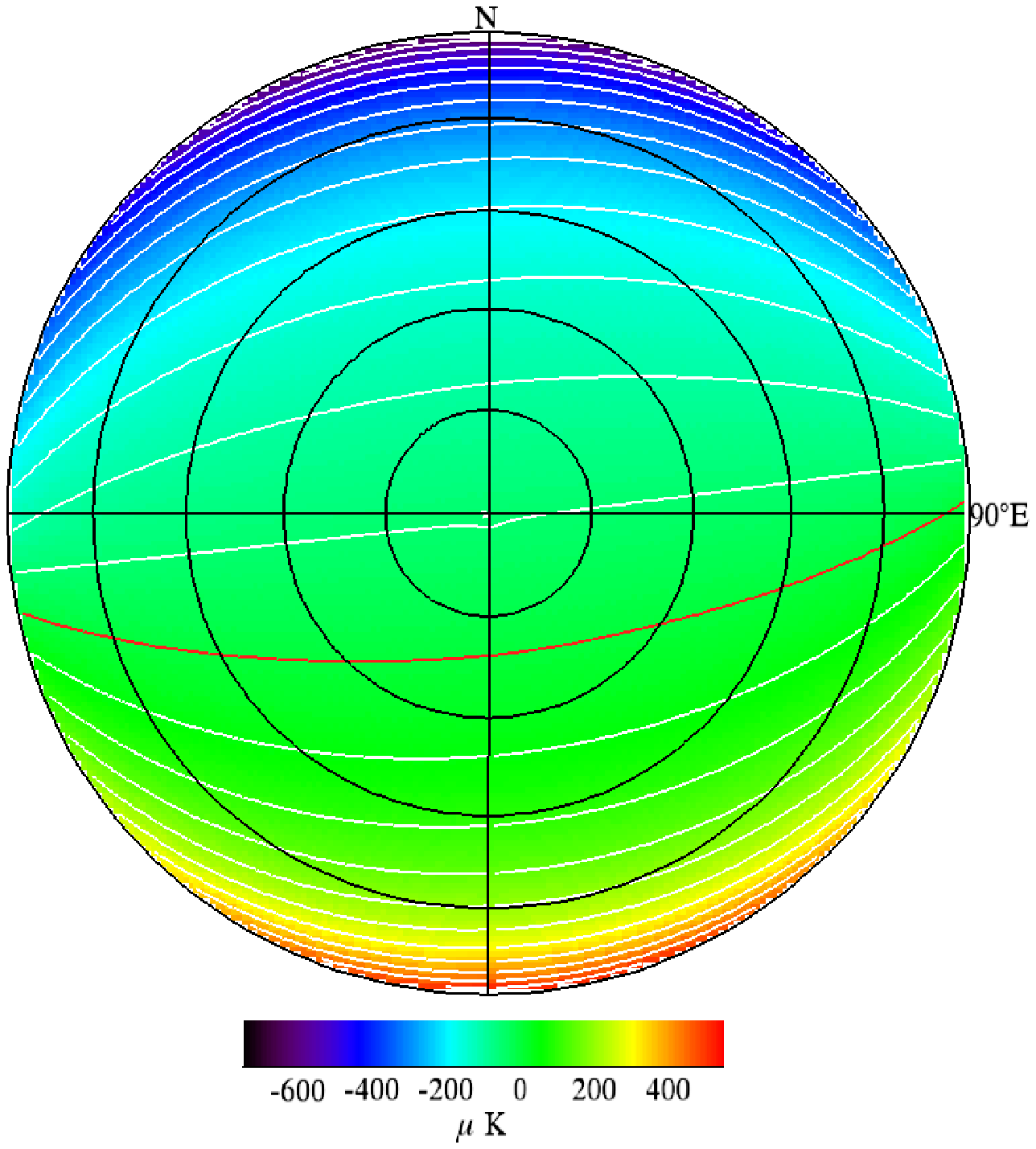}
\caption{Map of the atmospheric $O_2$ circularly polarized signal
at Atacama. The red contour denotes the null signal level. See
caption of fig. \ref{dc} for details.} \label{atacama}
\end{center}
\end{figure}

\begin{figure}
\begin{center}
\includegraphics[width=80mm]{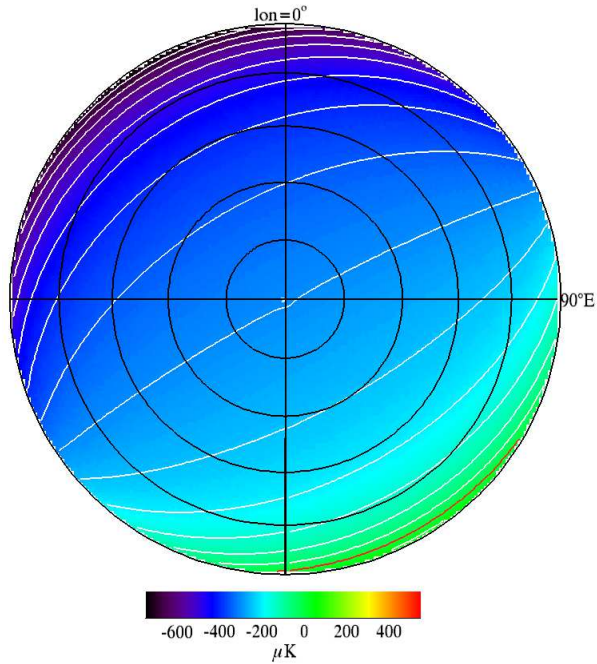}
\caption{Map of the atmospheric $O_2$ circularly polarized signal
at South Pole. See caption of fig. \ref{dc} for details, but
notice that upward direction is the geographic longitude $\ell =
0$.} \label{sp}
\end{center}
\end{figure}

\begin{figure}
\begin{center}
\includegraphics[width=80mm]{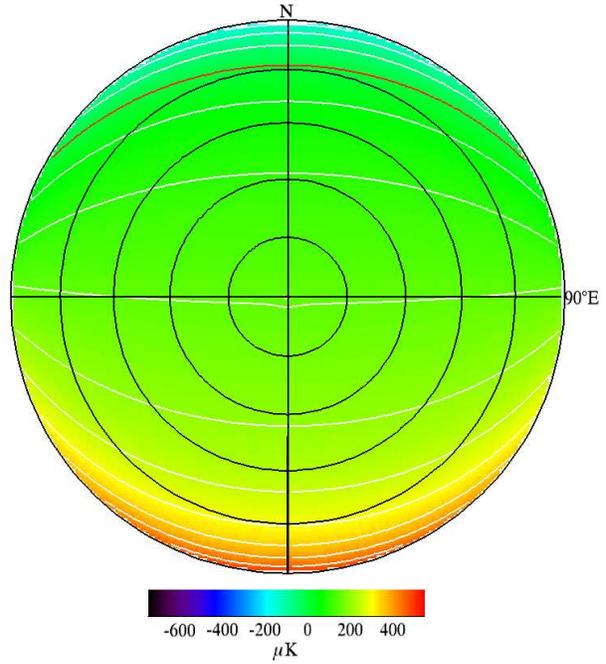}
\caption{Map of the atmospheric $O_2$ circularly polarized signal
at Testa Grigia. The red contour denotes the null signal level.
See caption of fig. \ref{dc} for details.} \label{tg}
\end{center}
\end{figure}

\begin{figure}
\begin{center}
\includegraphics[width=75mm]{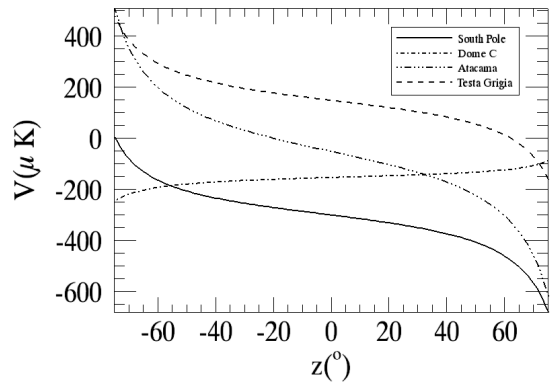}
\caption{Elevation scans of the Oxygen polarized signal for the
several sites. Cuts are taken along the North-South direction.
Frequency is $\nu = 90$ GHz.} \label{cut_elev_NS}
\end{center}
\end{figure}

\begin{figure}
\begin{center}
\includegraphics[width=75mm]{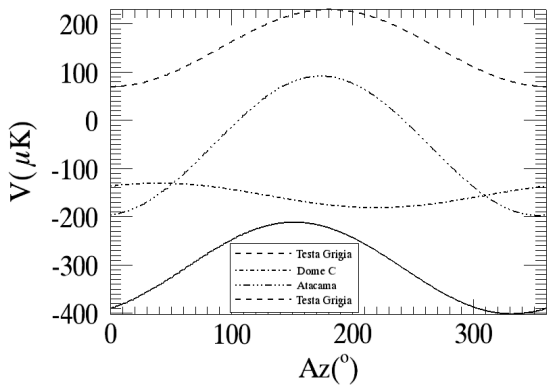}
\caption{Azimuth scan of the Oxygen polarized signal for the
several sites. Cuts are taken at constant elevation (45$^{\circ}$).
Frequency is $\nu = 90$ GHz.} \label{cut_az_45}
\end{center}
\end{figure}

We then built the maps of the polarized signal as seen at the
various sites in local alt-azimuthal coordinates. Maps of the
polarized signal, at 90 GHz, seen at the selected sites are shown
in Figures \ref{dc}, \ref{atacama}, \ref{sp} and \ref{tg}. Both
positive and negative values indicate that both right and left
handed circular polarization is present. We find $V=0$ in the maps
(red contour) at the angular positions where \textbf{B} is
orthogonal to the line of sight. Elevation cuts along the
North-South direction and azimuth cuts at constant elevation of 45$^{\circ}$ of these
maps are presented respectively in Figures
\ref{cut_elev_NS} and \ref{cut_az_45}. Observing in directions
different from the zenith the actual signal depends also on the
atmospheric thickness. This effect is visible in the elevation
scans that show a zenith-secant law dependence. The spatial
structure of these maps is dominated by the combination of
magnetic field direction and atmospheric thickness in the line of
sight. Each map covers a surface, which has a typical scale length
of $\sim 200$ km, at an altitude of $\sim 30$ km, corresponding to
a multipole $n \sim 200$. The magnetic field components at these
multipoles (or even higher) are negligible with respect to the
longer wavelength components and not larger than the accuracy of
the IGRF-2010 model ($\sim$ 10 nT).

The effect due to the variation of the atmospheric thickness is
combined in the maps with the direction of the magnetic field. At
Dome C, where the magnetic field is almost vertical, the two
effects compensate each other and the signal is flat on a large
part of the visible sky. Conversely at Atacama, where the magnetic field is
almost horizontal, the two effects combine increasing
the gradient along the visible sky.

The typical signal variation, at 90 GHz, for North-South elevation
scans (down to 60$^{\circ}$ from the zenith) is of the order of 70 $\mu
K$ at Dome C, of the order of 300 $\mu K$ at South Pole and Testa
Grigia, and of the order of 500 $\mu K$ at Atacama. The typical
signal variation for full-360$^{\circ}$ azimuth scans at $el=45^{\circ}$ is of the order of
50 $\mu K$ at Dome C, of the order of 170 $\mu K$ at South Pole
and Testa Grigia, and of the order of 270 $\mu K$ at Atacama.

\noindent

The signal gradient has also been computed in order to give a
better description of the signal angular dependence. Reference
values are summarized in Table \ref{tab_grad}. The column 3 of
Table \ref{tab_grad} presents the minimum and maximum gradients in
the elevation scans shown in Figure \ref{cut_elev_NS}, occurring
respectively at the zenith and at the lowest elevation. We
consider only the central part where $|el| \geq 45^{\circ}$. It
can be noticed that the signal variation is minimum ($\sim 0.3 -
0.8\mu K/^{\circ}$) at Dome C, while the gradient increases by a
factor of 4 at South Pole and Testa Grigia and by a factor of 7 at
Atacama. The column 4 of Table \ref{tab_grad} presents the maximum
signal gradient for the azimuth scans of Figure \ref{cut_az_45}.
These values indicate that Dome C presents the smoother pattern
for the polarized emission from atmospheric $O_2$.

\begin{table*}
\caption{Typical signals and gradients of the Stokes parameter $V$
at the several sites. Signals are evaluated at 90 GHz.
\emph{Column 2}: total polarized signal at the zenith (see Fig.
\ref{pol_comparison}); \emph{Column 3}: elevation gradient for a
North-South scan, minimum (occurring around the zenith) and
maximum (occurring al lowest elevation) values for $el \geq 45^{\circ}$
(see Fig. \ref{cut_elev_NS}); \emph{Column 4}: maximum azimuth
gradient for a scan at $el = 45^{\circ}$ (see Fig. \ref{cut_az_45});
\emph{Column 5}: peak-to-peak polarized signal in a scan at $el =
45^{\circ}$ (see Fig. \ref{cut_az_45}); \emph{Column 6}: azimuth
positions where maximum azimuth gradients occur (for azimuth
reference see Figs. \ref{dc} - \ref{tg}).}\label{tab_grad}
\begin{tabular}{|l|r|c|r|r|c|}
  \hline
  Site & $V_{z}$($\mu K$) & $\frac{\delta V}{\delta \theta_{el}}$($\mu K/^{\circ}$) &
  $\frac{\delta V}{\delta \phi_{az}}$($\mu K/^{\circ}$) &
  $\delta V_{pp}(az)$($\mu K$) & $\Phi_{az}$($^{\circ}$) \\
  \hline
  Dome C       & -153.4 \ & $0.3 - 0.8$ & 0.4 \ \ \ \ \ & 50.2  \ \ \ \ \ & 129; 310  \\
  South Pole   & -299.0 \ & $1.4 - 3.2$ & 1.6  \ \ \ \ \ & 188.1 \ \ \ \ \ & \ \ 61; 242 \\
  Atacama      &  -52.5 \ & $2.5 - 5.1$ & 2.5  \ \ \ \ \ & 288.0 \ \ \ \ \ & \ \ 83; 264 \\
  Testa Grigia & +147.1 \ & $1.3 - 2.9$ & 1.4  \ \ \ \ \ & 158.8 \ \ \ \ \ & \ \ 90; 271 \\
  \hline
\end{tabular}
\end{table*}

By means of our template the $O_2$ polarized signal can be
computed at any time given the site where the observations are
carried out, giving the possibility of subtracting it from the sky
signal. For a real experimental setup the signals computed for a
given direction have to be integrated with the efficiency of the
bandwidth and convolved with the observing beam profile.

\subsection{Bands optimization}
\begin{figure}
\includegraphics[width=75mm]{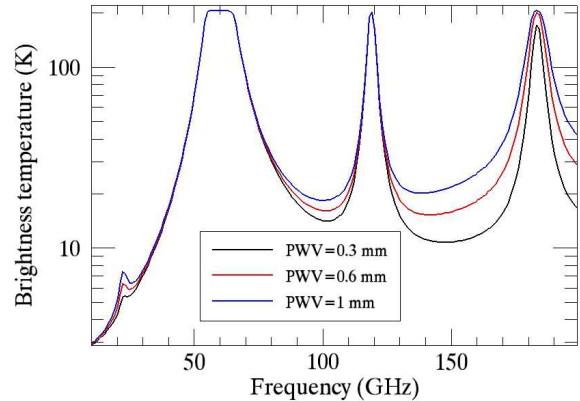}
\caption{Brightness temperature of the atmosphere at the zenith
for Dome C station. The three curves correspond to different
values of the precipitable water vapour (PWV=1 mm corresponding to
a typical summer condition, PWV=0.6 mm and PWV=0.3 mm
corresponding to a typical winter condition).}
\label{unpol_opacity}
\end{figure}

Astrophysical and cosmological signals can be investigated by
ground based experiments through the atmospheric windows, where
transparency is large. The opacity of the atmosphere at mm-wave
frequency is due to the combined unpolarized emission of (mainly)
molecular Oxygen (the same transitions considered for the
polarized emission) and water vapor lines (at 22 GHz and 183 GHz).
In Fig. \ref{unpol_opacity} the total brightness temperature of
the atmosphere at Dome C is shown between 10 GHz and 200 GHz.

\begin{figure}
\begin{center}
\includegraphics[width=75mm]{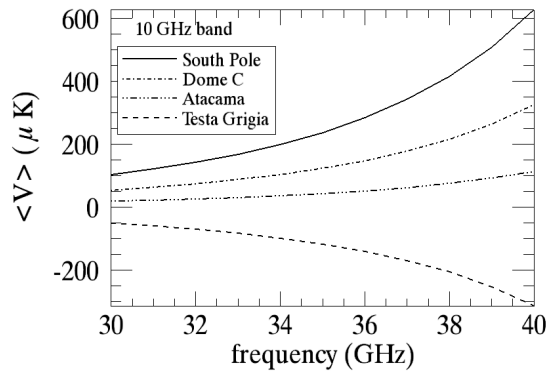}
\caption{Integrated polarized signal from $O_2$ vs the central
frequency: bandwidth $\Delta \nu = 10$ GHz; frequency interval
from 25 to 45 GHz; observing towards the zenith.} \label{tbv26-60}
\end{center}
\end{figure}

\begin{figure}
\begin{center}
\includegraphics[width=75mm]{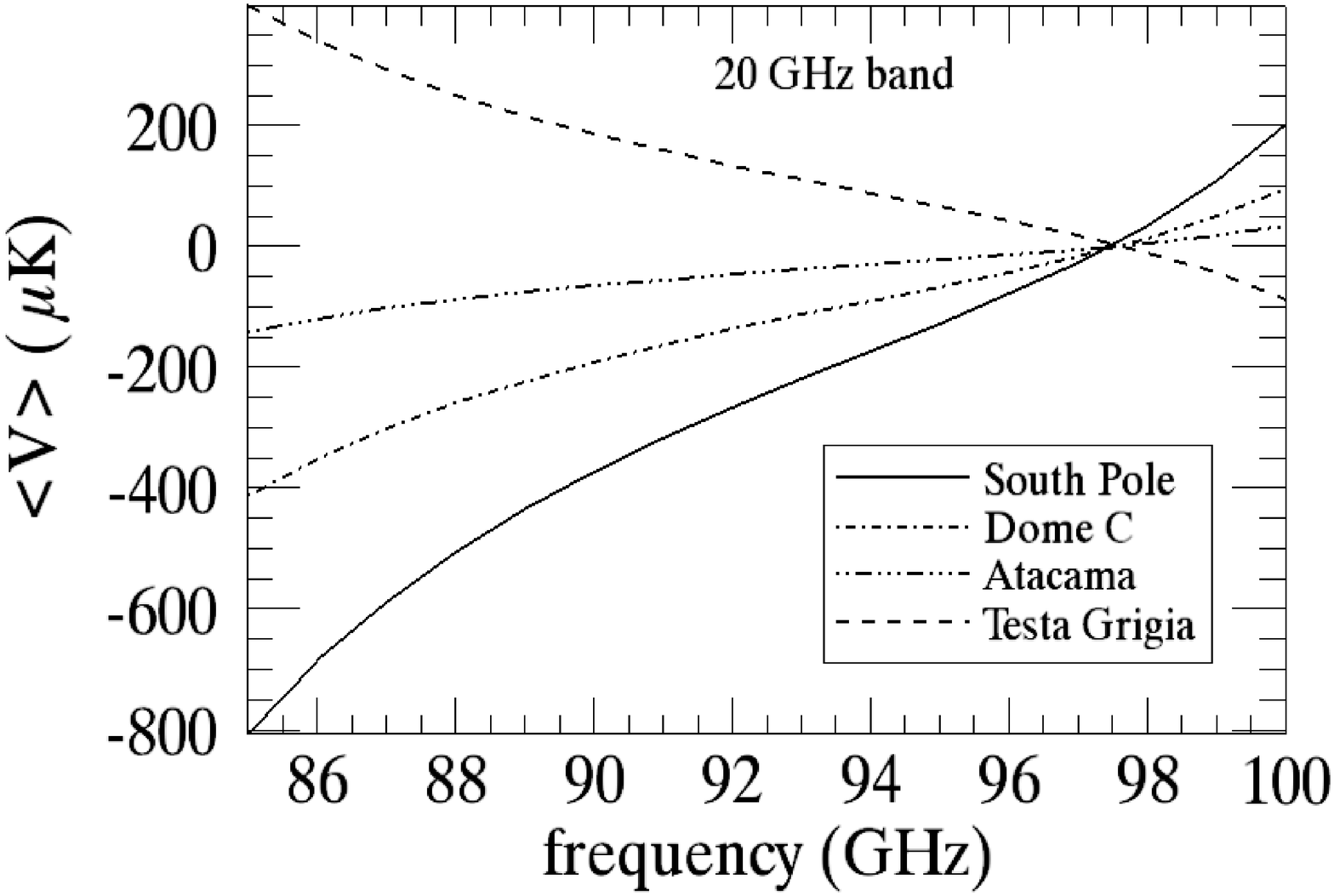}
\caption{Integrated polarized signal from $O_2$ vs the central
frequency: bandwidth $\Delta \nu = 20$ GHz; frequency interval
from 75 to 110 GHz; observing towards the zenith.}
\label{tbv75-110}
\end{center}
\end{figure}

\begin{figure}
\begin{center}
\includegraphics[width=75mm]{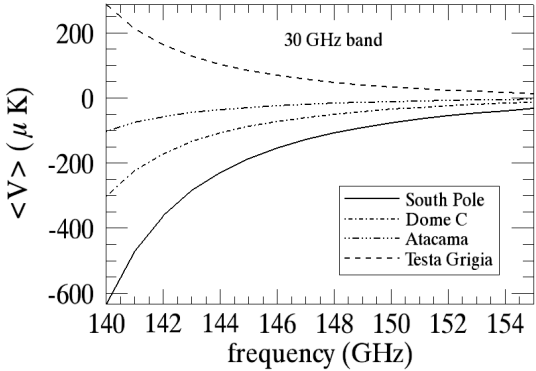}
\caption{Integrated polarized signal from $O_2$ vs the central
frequency: bandwidth $\Delta \nu = 30$ GHz; frequency interval
from 125 to 170 GHz; observing towards the zenith.}
\label{tbv120-170}
\end{center}
\end{figure}

Inside this interval we define three frequency windows of
atmospheric transparency: the first between 25 GHz and 45 GHz, the
second between 75 GHz and 110 GHz and the third between 125 GHz
and 170 GHz. Inside these frequency intervals it is possible to
observe the sky with a transparency larger than 0.8, using a wide
bandwidth ($\Delta\nu/\nu=0.2-0.3$). In these frequency bands we
estimate the integrated polarized temperature due to atmospheric
oxygen. We select 3 ideal rectangular bandpass which are 10 GHz,
20 GHz and 30 GHz wide (i.e. corresponding roughly to a 20\%
bandpass) respectively for the three bands. The results are shown
in Figures \ref{tbv26-60}, \ref{tbv75-110} and \ref{tbv120-170} as
a function of the central observing frequency.

This analysis allows us to define the optimal frequency bandwidths
for observation of circular polarization inside the windows, while
avoiding the largest signals from the peaks and the highest part
of the line wings. In the 75-110 GHz interval the high frequency
side is preferred because here we find a null in the polarized
signal. Therefore the optimal frequency band is 87-107 GHz. Inside
the 125-170 GHz interval we do not find a null, but rather a signal
that becomes smaller and smaller as we rise the central frequency of
our integration band: for this reason, the contamination is
smaller in the 140-170 GHz interval.
Also in the 25-45 GHz interval there is not a null, therefore the
farther from the line tails the smaller the polarized
contribution, and the low frequency side is preferred (25-35 GHz).
Anyway, it has to be kept in mind that a global optimization of
the bandwidth must also consider the unpolarized signal (see Fig.
\ref{unpol_opacity}) and the true bandwidth transmission.

\subsection{Accuracy of the templates}\label{acc_temp}
We have estimated the accuracy of the templates at 90 GHz for the various
sites, taking into account uncertainties and typical variability
of the principal parameters of the model, i.e. the magnetic field
and the atmospheric conditions. We first considered the magnetic
field$^4$
and summarize the results in Table \ref{tab_Err_B}, taking into
account several sources of uncertainty or variations: \textbf{1)}
We estimated a typical rms uncertainty: $\delta |B|$=20nT, $\delta
\hat{p} $=$\delta \hat{q}$=1$^{\prime}$. But at Dome C and South
Pole we consider $\delta \hat{q}$=5$^{\prime}$. This uncertainty
includes both the quoted error bars of the IGRF model and the
unaccounted contribution of the spatial frequencies higher than
$n=13$. Results, obtained using a Monte Carlo for these
parameters, are reported in column 3 of Table \ref{tab_Err_B}.
\textbf{2)} Then we evaluated the secular variation (SV) of the
magnetic field, for the various sites. We considered the
consolidated values for the past since 2005 and the forecast
values for the next years up to 2015. We found a nearly linear
variation with time, which we reported in column 4 of Table
\ref{tab_Err_B}. Using the magnetic field values obtained for the
several years we computed the amplitude of the atmospheric
circular polarization. We found again a nearly linear variation
with time, reported in column 5 of Table \ref{tab_Err_B}. At
Atacama the change of magnetic field direction plays an important
role, in particular at the zenith, while in all the other sites
the strength variation is dominating. \textbf{3)} Finally we
evaluated the variation of the atmospheric circular polarization
in case of an intense magnetic storm when the magnetic field
strength changes as much as $\sim$ 1000 nT. Usually magnetic
disturbances last no more than a few days and afterwords quiet
conditions are established again. Results, which represent unusual
maximum short-time variations, are reported in column 6 of Table
\ref{tab_Err_B}.

\begin{table*}
\caption{Typical polarized signal uncertainties or variations due
to the amplitude od the Earth magnetic field. We evaluated the
effect at the zenith and at ($el$=$45^{\circ}$ ; $az$=0).
\emph{Column 3}: rms error bar due to the IGRF model
uncertainties. \emph{Column 4}: Secular variation of the magnetic
field. \emph{Column 5}: change of the signal due to SV of magnetic
field. \emph{Column 6}: signal variation due to a typical big
magnetic disturbance (md).}\label{tab_Err_B}
\begin{tabular}{|l|c|c|c|c|r|}
  \hline
  Site & Elevation & $\delta V_{rms}$($\mu K$) &
  $\frac{\delta |B|_{sv}}{\delta t}$($nT/y$) &
  $\frac{\delta V_{sv}}{\delta t}$($\mu K/y$) & $\delta V_{md}$($\mu K$) \\
  \hline
  Dome C       & 90$^{\circ}$ & $0.05$ & -21 & $+0.05$ & $\sim$ 3  \ \ \ \ \ \\
  Dome C       & 45$^{\circ}$ & $0.07$ & -21 & $+0.13$ & $\sim$ 2  \ \ \ \ \ \\
  South Pole   & 90$^{\circ}$ & $0.11$ & -63 & $+0.38$ & $\sim$ 5  \ \ \ \ \ \\
  South Pole   & 45$^{\circ}$ & $0.16$ & -63 & $+0.44$ & $\sim$ 7 \ \ \ \ \ \\
  Atacama      & 90$^{\circ}$ & $0.06$ & -52 & $-0.27$ & $\sim$ 2  \ \ \ \ \ \\
  Atacama      & 45$^{\circ}$ & $0.17$ & -52 & $+0.23$ & $\sim$ 9 \ \ \ \ \ \\
  Testa Grigia & 90$^{\circ}$ & $0.07$ & +26 & $+0.12$ & $\sim$ 3  \ \ \ \ \ \\
  Testa Grigia & 45$^{\circ}$ & $0.07$ & +26 & $+0.04$ & $\sim$ 1  \ \ \ \ \ \\
  \hline
\end{tabular}
\end{table*}

We then considered the atmospheric parameters and summarize the
results in Table \ref{tab_Err_TP}. We limited our analysis to
Atacama$^1$
Dome C\footnote{http:$//$www.climantartide.it} and South Pole$^2$
because of the poor statistics available for Testa Grigia. We took
into account several sources of uncertainty or variations:
\textbf{1)} We estimated the uncertainty due to the accuracy of
the measurements in the temperature and pressure profile. We took
the values quoted by \citet{tomasi} ($\delta P_{exp}=5$ mb,
$\delta T_{exp}=0.5$ K). Results ($\delta V_{exp}$), obtained
using a Monte Carlo distribution for these parameters, are
reported in column 3 of Table \ref{tab_Err_TP}. \textbf{2)} We
then estimated the variability of $T$ and $P$ comparing several
measured profiles, taken in different days, but equivalent
conditions. We found a nearly equivalent rms variation for the
various sites: $\delta P_{st} \simeq 0.8$ mb, $\delta T_{st}
\simeq 2.0$ K. Using these variabilities we obtain $\delta
V_{st}$, the signal variation reported in column 4 of Table
\ref{tab_Err_TP}. We searched also for a day-night variation, but
we found that it is limited to the temperature of the lowest
layers of the atmosphere. The effect we found is not larger than
$\delta V_{st}$. \textbf{3)} We also estimated the seasonal
long-term variability of the atmospheric parameters. We compared
measured profiles taken several months later and evaluated the
maximum variation of $T$ and $P$. We found that at Atacama the
variation, even if relevant, is still limited to the lowest layers
of the atmosphere. Conversely at Dome C and South Pole the
seasonal variation impacts the full air column. In columns 5 and 6
of Table \ref{tab_Err_TP} we report the typical maximum seasonal
variation measured in the lowest layers of the atmosphere (the
maximum temperature variation at the ground is usually even
larger). Results on the polarized signal ($\delta V_{lt}$) are
reported in column 7 of Table \ref{tab_Err_TP}.

\begin{table*}
\caption{Typical polarized signal uncertainties or variations due
to the atmospheric parameters. We evaluated the effect at the
zenith and at ($el$=$45^{\circ}$ ; $az$=0). \emph{Column 3}: rms
error bar due to temperature and pressure profile measurement
uncertainties ($\delta P=5$ mb, $\delta T=0.5$ K). \emph{Column
4}: signal variation due to short-term rms variation of
temperature and pressure. \emph{Column 5}: typical seasonal
(long-term) temperature variation. \emph{Column 6}: typical
seasonal (long-term) pressure variation. \emph{Column 7}: signal
variation due to a typical seasonal (long-term)
variation.}\label{tab_Err_TP}
\begin{tabular}{|l|c|c|c|c|c|c|}
  \hline
  Site & Elevation & $\delta V_{exp}(\mu K)$ &
  $\delta V_{st}(\mu K)$ & $\delta T_{lt}(K)$ &
  $\delta P_{lt}(mb)$ & $\delta V_{lt}(\mu K)$ \\
  \hline
  Dome C       & 90$^{\circ}$ & 0.7 & 0.5  & 36   & 15   &  3.4  \\
  Dome C       & 45$^{\circ}$ & 0.7 & 0.5  & 36   & 15   &  2.9  \\
  South Pole   & 90$^{\circ}$ & 1.1 & 0.8  & 26   & 15   & 10.0  \\
  South Pole   & 45$^{\circ}$ & 1.4 & 1.1  & 26   & 15   & 13.7  \\
  Atacama      & 90$^{\circ}$ & 0.1 & 0.1 & 10.5 &  2.6 &  0.23 \\
  Atacama      & 45$^{\circ}$ & 0.6 & 0.3  & 10.5 &  2.6 &  0.9  \\
  \hline
\end{tabular}
\end{table*}

As shown by Tables \ref{tab_Err_B} and \ref{tab_Err_TP} the
accuracy is of the order of $\mu$K. Larger effects arise in case
of large magnetic disturbances or seasonal changes. We did not
take into account the line profile model. The line frequencies are
known with negligible error bars, but the Oxigen absorption values
have error bars of at least 5\% at 90 GHz. This accuracy applies
up to $\sim$ 120 GHz, but at higher frequencies model and values
of Oxigen absorption are not yet validated and calculations should
be used with caution. The effect of these uncertainties should be
added to the accuracy quoted previously. It is also important to
stress that the computed uncertainty refers to the total power
polarized signal. Usually we are interested to subtract signal
gradients between different observing directions, and the related
uncertainty is definitely smaller than the total estimated value
of $\delta V$, in particular for small angular scales.

\section{Conclusions}
In this paper we evaluate the contribution of atmospheric $O_2$
Zeeman split roto-vibrational lines in the mm-wave part of the
electromagnetic spectrum, seen as a contaminant for CMB
polarization experiments. A code to simulate this signal for
various sites (Dome C and South Pole in the Antarctica, Testa
Grigia in the Italian Alps and Atacama, Chile) of interest for
mm-wave astronomy has been developed on the basis of the theory
developed in \cite{len2} and \cite{len1}). In this way we compute
a template of the atmospheric circularly polarized signal for
every site provided that the vertical profiles of atmospheric
pressure and temperature are available.

Maps of the circularly polarized signal in local coordinates have
been obtained for each site examined. These maps show a
dipole-like large scale feature with a maximum gradient aligned
approximately, for low-mid latitude sites, with the North-South
direction, and for polar sites, along the magnetic field lines.
Atacama is the examined place where the polarized signal at the
zenith is the lowest. This is due to the altitude of the site and
to the amplitude and direction of the magnetic field.
Unfortunately scanning experiments at Atacama suffer a larger
polarized gradient with respect to other sites. In fact the
angular variation of the signal is of order $\sim 1.3-5.1 \ \mu
K/^{\circ}$ at 90 GHz for all sites except Dome C, where it is a factor
$\sim$ 4$-$7 lower, because here the atmospheric polarized signal is
the flattest, along the full local visible sky.

We then evaluated the integrated polarized signal inside an ideal
bandpass of a wide-band ($\Delta\nu/\nu=0.2$) experiment for 3
mm-wave atmospheric windows between 10 and 200 GHz. We found the
frequencies where it is more convenient to observe, because here
the signal vanishes or decreases significantly. Finally, an estimate of the template accuracy
and variation has been done, using the main features of the
magnetic field and the statistics of measured atmospheric
profiles. At 90 GHz, we obtained a typical uncertainty of the order of $\sim$
1 $\mu$K, dominated by accuracy and variability of the
atmospheric parameters. Large variations (up to $\sim$ 10-15
$\mu$K) are expected in case of large magnetic disturbances and
due to seasonal atmospheric variations. Uncertainty and secular
variation of the magnetic field have, on the other hand, a
negligible effect.

A linear polarization is also generated and it is larger when the
line of sight is orthogonal to the Earth magnetic field. However
the ratio of linear to circular polarization is $\sim 10^{-4}$,
therefore still below the current detection sensitivity. Moreover
the non vanishing component of linear polarization is nearly
aligned with the Earth magnetic field.

\section*{Acnowledgments}
Authors want to acknowledge P. Rosenkranz and S. Hanany for the
precious contribution to the improvement of the paper. Data from
Dome C are Copyright of Italian PNRA (http:$//$www.climantartide.it).

\end{document}